\documentclass[%
reprint,  
superscriptaddress,
nofootinbib,
amsmath,amssymb,
aps,
pra,
raggedbottom,
onecolumn,
xcolor,
]{revtex4-1}

\usepackage{graphicx}% Include figure files
\usepackage{dcolumn}% Align table columns on decimal point
\usepackage{bm}% bold math
\usepackage{lipsum}
\usepackage{float}
\usepackage{subfigure}
\usepackage{natbib}
\setcitestyle{super}
\usepackage{filecontents}
\usepackage[colorlinks=true,bookmarks=false]{hyperref}
\hypersetup{linkcolor=blue,citecolor=red,urlcolor=blue}   %\usepackage[mathlines]{lineno}% Enable numbering of text and display math
\begin{document}

%\preprint{APS/123-QED}

\title{Self-charging of identical grains in the absence of an external field}% Force line breaks with \\

\author{R. Yoshimatsu} \email{ryutay@phys.ethz.ch} 
\affiliation{%
Computational Physics for Engineering Materials, IfB, ETH Zurich, Wolfgang-Pauli-Strasse 27, 8093 Zurich, Switzerland
}

\author{N.A.M. Ara\'ujo}
\affiliation{%
Departamento de F\'isica, Faculdade de Ci\v encias, Universidade de Lisboa, P-1749-016 Lisboa, Portugal, and Centro de F\'isica Te\'orica e Computacional, Universidade de Lisboa, P-1749-016 Lisboa, Portugal
}

\author{G. Wurm}
\affiliation{%
Faculty of Physics, University of Duisburg-Essen, Lotharstr. 1, D-47057 Duisburg, Germany
}

\author{H.J. Herrmann}
\affiliation{%
Computational Physics for Engineering Materials, IfB, ETH Zurich, Wolfgang-Pauli-Strasse 27, 8093 Zurich, Switzerland
}
\affiliation{%
Departamento de F\'isica, Universidade Federal do Cear\'a, 60451-970 Fortaleza,Cear\'a, Brazil
}

\author{T. Shinbrot}  \email{shinbrot@rutgers.edu } 
\affiliation{%
Computational Physics for Engineering Materials, IfB, ETH Zurich, Wolfgang-Pauli-Strasse 27, 8093 Zurich, Switzerland
}
\affiliation{%
Department of Biomedical Engineering, Rutgers University, Piscataway, New Jersey, 08854, USA 
}

%\altaffiliation[Also at]{Physics Department, XYZ University.}%Lines break automatically or can be forced with \\
%\author{Author 2}%
% \affiliation{%
%  Department of Physics ETH-Z\"{u}rich
% }%
                  
\date{September 2, 2016}
\begin{abstract}
We investigate the electrostatic charging of an agitated bed of identical grains using simulations, mathematical modeling, and experiments. We simulate charging with a discrete-element model including electrical multipoles and find that infinitesimally small initial charges can grow exponentially rapidly. We propose a mathematical Turing model that defines conditions for exponential charging to occur and provides insights into the mechanisms involved. Finally, we confirm the predicted exponential growth in experiments using vibrated grains under microgravity, and we describe novel predicted spatiotemporal states that merit further study.
\end{abstract}

\maketitle

%\tableofcontents
\section*{\label{introduction}Introduction}
In 1963, the volcanic island Surtsey, named after the legendary fire giant Surtr, rose out of the North Atlantic Ocean. True to its name, over the next year and a half, the island's volcanic debris cloud spat fire in the form of multimillion volt lightning displays \cite{anderson1965electricity}. Desert sandstorms similarly have long been known to generate lightning \cite{shaw1929tribo}.

How grains generate charge in volcanic plumes, sandstorms Ð or industrial problems such as pharmaceutical mixing \cite{mehrotra2007spontaneous} or dust explosions \cite{hertzberg1987introduction} Ð remains controversial \cite{schein2007recent}. Mechanisms have been proposed for charging granular materials that differ in shape, size, or chemical composition \cite{lowell1986triboelectrification, terris1989contact, forward2009charge, kok2009electrification, waitukaitis2014size, cademartiri2012simple, kolehmainen2016hybrid, harper1998contact}. However, charging is also observed for materials that are absolutely identical, and experiments show that charging grows with repeated contacts \cite{shinbrot2008spontaneous, baytekin2011mosaic}. Beyond the unexpected nature of these findings, the fundamentally surprising thing about charging of identical materials is that one appears to get something from nothing: charge that requires energy appears from materials that ought to discharge one another on contact. 

Previous work has shown that one origin of this energy can be an external electric field that feeds electrification \cite{pahtz2010particle, C6SM00357E, zheng2004theoretical, rasmussen2009enhancement, zhang2015electric}. In the present work, we demonstrate that, remarkably, an external field is not needed: infinitesimal charges on grains themselves can induce charges on their neighbors, bootstrapping one another to grow exponentially rapidly in agitated beds. Unlike prior work \cite{siu2014self}, we show that the energy for this charge growth can arise strictly conservatively, trading mechanical work for electrical energy. It is by no means self-evident whether or when this could cause a buildup of significant charges, but as we will show, this feedback mechanism can promote a rich set of macroscopic charge patterns within the bed and can generate very large electric fields.
\begin{figure}[htb]
\centering
\includegraphics[width=9cm]{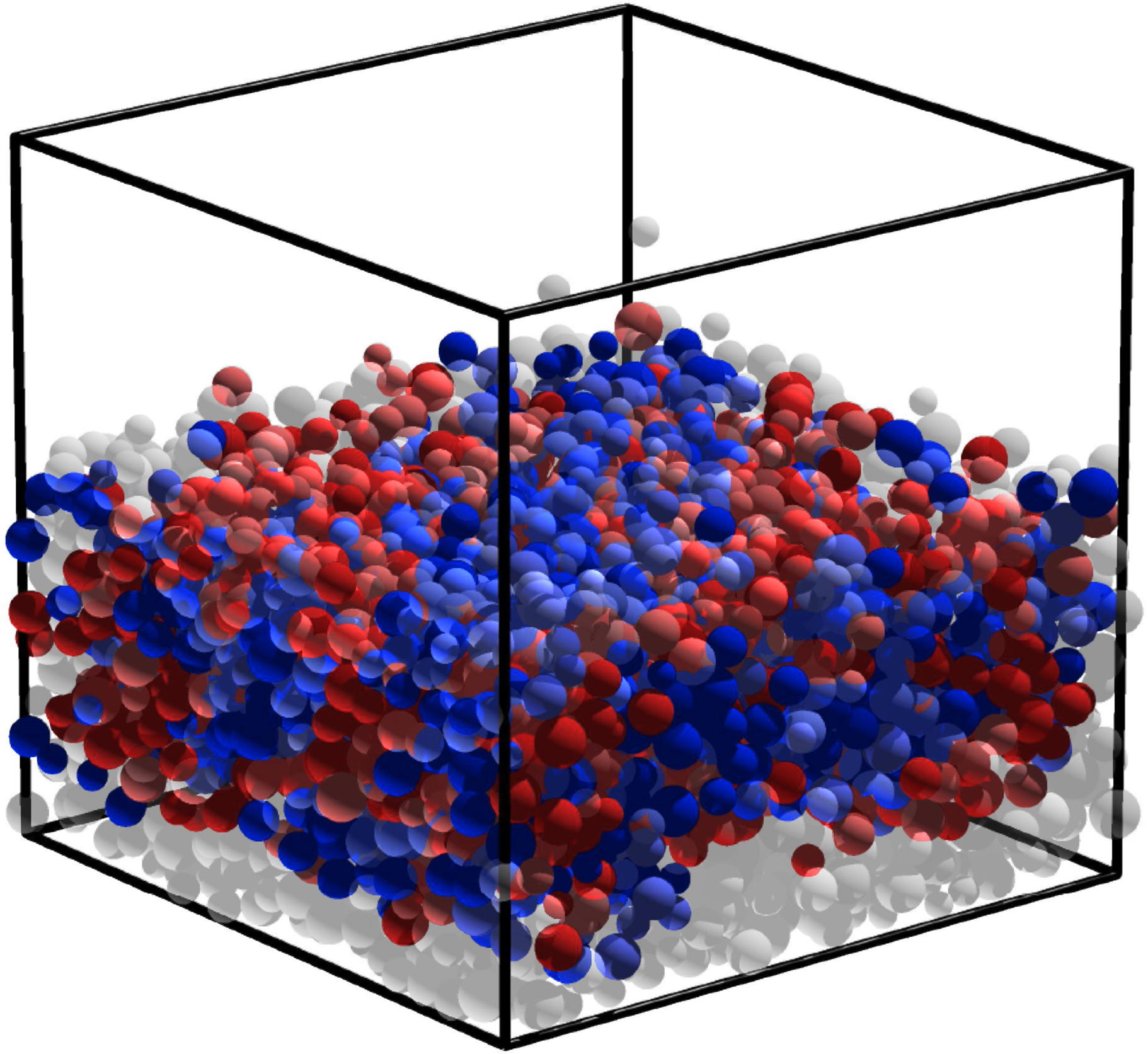}
\caption{Typical snapshot of simulation showing the charging of an agitated granular bed. Here we use $\chi_e$=$\eta$=1 as defined in text, and show time after 5 seconds. Grains are colored according to their net charge defined as: $\sum_{k=1}^{6}q_{i,k}$ (blue: negative, red: positive); with grains with an absolute net charge magnitude below $20$ pC in transparent grey.}
\label{profile}
\end{figure}

The underlying mechanism that we explore here is that the electric field from charged particles induces a polarization on a neighbor given by,
\begin{equation}
\vec{p}^{\ \mathrm{ind}}_{i}=\frac{\chi_{e} R_{d}^{3} }{k_{e}}\vec{E}_{i}.
\label{charge_induced}
\end{equation}
where $\chi_{e}$ is the grain polarizability ranging from zero to unity, $R_{d}$ is an effective dipole radius, $\vec{E}_i$ is the electric field at the center of the $i^{th}$ grain due to surrounding permanent charges, and $k_e$ is Coulomb's constant. In a bed of colliding grains, we also assume that when the $i^{th}$ and the $j^{th}$ particles come into contact, charges $q_{i,k}$ and $q_{j,l}$ on the grain surfaces can partially neutralize \cite{pahtz2010particle, C6SM00357E} according to:
\begin{align}
\begin{split}
q'_{i,k}=\left(1-\frac{\eta}{2}\right)q_{i,k}+\frac{\eta}{2}q_{j,l},\\
q'_{j,l}=\left(1-\frac{\eta}{2}\right)q_{j,l}+\frac{\eta}{2}q_{i,k},
\end{split}
\label{charge_transfer}
\end{align} 
where the prime denotes the constituent charges after collision, $\eta$ is a neutralization efficiency, and $k$ and $l$ denote the two constituent charges of the $i^{th}$ and $j^{th}$ grains, respectively, that are in contact upon collision. We assume that particles initially have charges of the order of 10$^5$ elementary charges and we observe a growth by three orders of magnitude as we will show later. Details of the mechanism defined by Eqs.\,\eqref{charge_induced} and \,\eqref{charge_transfer} appear in Methods.

\section*{\label{results}Results}
We describe a particle-based simulation of charging of agitated grains subject to Eqs.\,\eqref{charge_induced} and \,\eqref{charge_transfer} followed by a continuum mathematical model, with the goal of establishing whether induced polarization combined with contact neutralization can predictably amplify small initial charges. We conclude with an experiment confirming that this amplification occurs as predicted. 

\subsection*{Discrete element simulation}
We begin by performing discrete-element method (DEM) \cite{granular} simulations of mechanical and electrical interactions between insulating particles. Typical qualitative charges on grains are shown in Fig.\,\ref{profile}, and details are presented in Methods. In summary, we use standard procedures to evaluate forces and torques on particles including both mechanical and electrostatic interactions for particles subject to Eqs.\,\eqref{charge_induced} and \,\eqref{charge_transfer}. We calculate polarizations and mechanical interactions by embedding within each particle three pairs of orthogonally placed constituent charges (see Methods, Fig.\,\ref{particle}) at fixed separations 2$R_d$, where $R_d$ is 2/3 of the mean grain radius. These constituent charges take part in neutralization events defined by Eq.\,\eqref{charge_transfer}. Our simulations are 3D and use horizontally periodic boundary conditions and a bottom surface that injects energy by vertically kicking impinging particles. For every grain, we set the initial charge of each of its domains randomly chosen from a uniform distribution between [-10$^{-2}$, 10$^{-2}$] pC.

We enforce energy conservation to compensate for both the work needed to separate induced charges within each particle and the energy associated with a dipole inducing secondary dipole moments on its neighbors. For the first energy compensation term, we apply a force on particles such that the work done by the force compensates exactly for the energy necessary to induce the corresponding dipoles, and for the second, we reduce dipole moments to account for the energy associated with surrounding field effects. Both terms are detailed in Methods. 

\begin{figure}[htb]
\centering
	\includegraphics[scale=1.5]{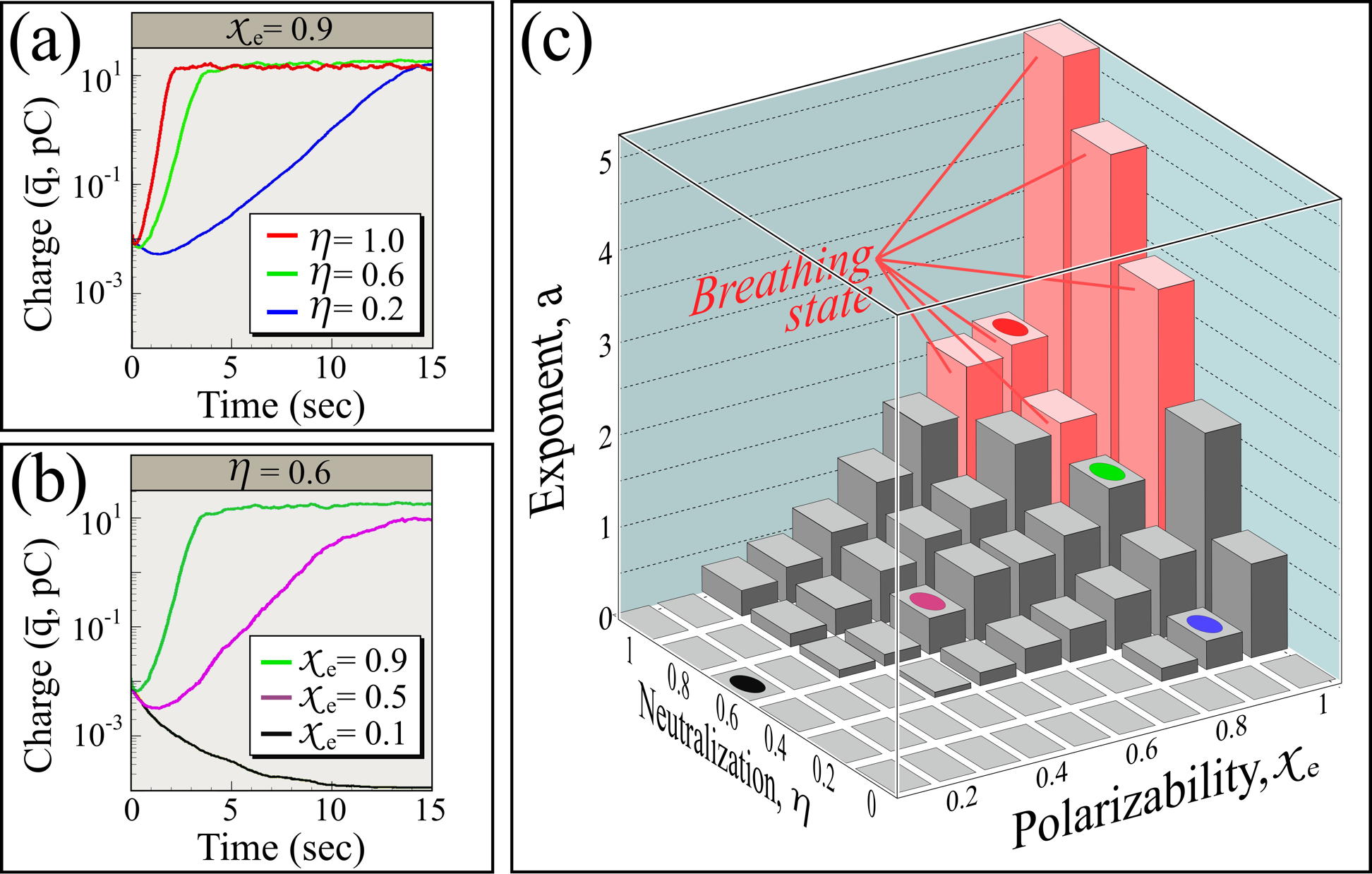}
	\caption{Time evolution of mean charge amplitude, $\bar{q}$, in log-linear plots. (a) $\chi_e$ is fixed and $\eta$ is varied. (b) $\eta$ is fixed and $\chi_e$ is varied. (c) Exponential growth rates from the slope of linear fits of log$_{10}$($\bar{q}$) vs. time plots for an for array of parameters. Growth rates are defined as: $a=\frac{1}{t}log_{10}\left(\frac{\bar{q}(t)}{b}\right)$, where $b$ is a constant. Colored spots correspond to colors used in panels (a) and (b), and breathing modes are discussed in text. }
\label{table_of_exponent}
\end{figure}

To quantify charge growth, we evaluate the evolution of the absolute value of all charges averaged over all grains: 
\begin{equation}
\bar{q}=\frac{1}{N}\sum_{i=1}^{N}\left| \sum_{k=1}^{6}q_{i,k} \right|,
\label{}
\end{equation}
as a function of model parameters $\chi_e$ and $\eta$, as shown in Fig.\,\ref{table_of_exponent} using $N=10^{3}$ particles (see Methods for initialization procedure). 

Figure\,\ref{table_of_exponent}(a) shows that for large constant polarizability, $\chi_{e}$, $\bar{q}$ typically does grow roughly exponentially following an initial transient and continuing up to an asymptote that we discuss shortly. We plot $\bar{q}$ here, but remark that polarizations, and charges of both signs, also grow with the same exponential rate. For fixed neutralization, $\eta$, $\bar{q}$ also exhibits an exponential growth period, however for small $\chi_{e}$, $\bar{q}$ decreases in time, as shown in Fig.\,\ref{table_of_exponent}(b). Figure\,\ref{table_of_exponent}(c) collects growth rates obtained from the slopes of least-squares fits in linear regions of log$_{10}(\bar{q})$ vs. time plots for $\chi_e$ and $\eta$ ranging from zero to one. Evidently, the growth rate increases with both $\chi_e$ and $\eta$. 

Figure\,\ref{table_of_exponent} contains several features that we discuss next. First, exponential growth is only seen for sufficiently large $\chi_e$ and $\eta$: Fig.\,\ref{table_of_exponent}(c) suggests that there is an onset criterion for exponential growth. Below this onset, $\bar{q}$ shrinks monotonically. Second, for most parameter values, exponential growth is preceded by a transient during which $\bar{q}$ drops. Third, $\bar{q}$ reaches an asymptotic value for long times. And fourth, Fig.\,\ref{table_of_exponent}(c) indicates an oscillatory ``breathing'' state that we will describe.

\begin{figure}[htb]
\centering
	\includegraphics[scale=1.75]{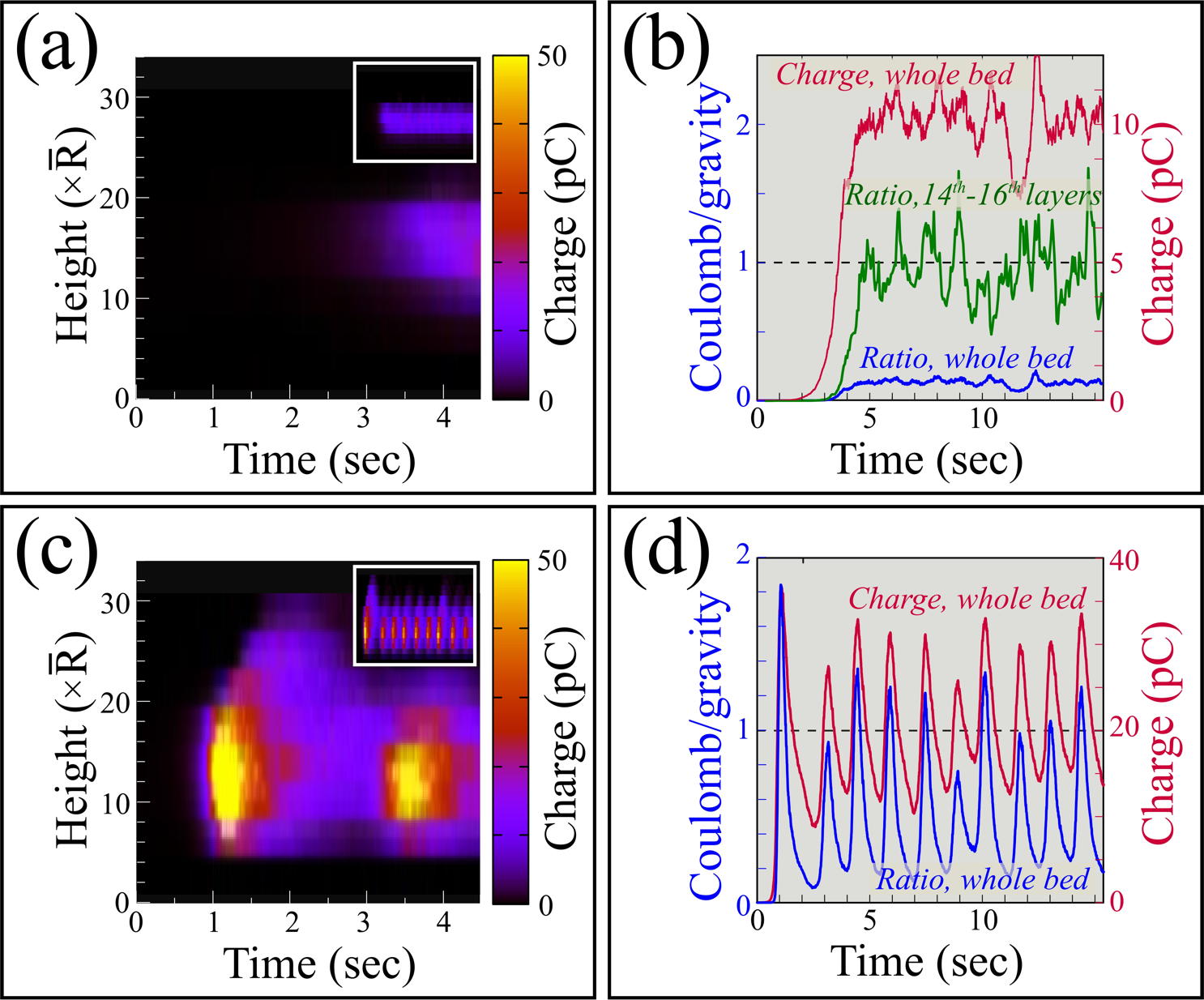}
	\caption{Spatiotemporal evolution of $\bar{q}$ and ratio between the magnitude of typical Coulomb and gravitational forces, $R_{cg}$. This ratio is given in cgs units \cite{kolehmainen2016hybrid} by $R_{cg}=\bar{q}^{2}/\bar{r}_{min}^{2}\bar{m}g$, where $\bar{r}_{min}$ is the mean of the distance to each grainÕs nearest neighbor, $\bar{m}$ is the mean grain mass, and $g$ is gravity. (a) Spatiotemporal evolution of bed charges using $\chi_e$=0.6 and $\eta$=1. Inset shows the same plot over longer time, (to 14 sec.). (b) Time evolution of $\bar{q}$ and $R_{cg}$ (blue: averaged over the entire bed, green: averaged over the layers 14-16), using $\chi_e$=0.6 and $\eta$=1. Note that although $R_{cg}$ is only 20\% when averaged over the entire bed, in the fastest charging region, around height=15, $R_{cg}$=1. (c) Spatiotemporal evolution of $\bar{q}$ using $\chi_e$=$\eta$=1: again inset shows same plot over longer time, (to 14 sec.), highlighting periodic oscillations in charge. Notice that for $\chi_e$=$\eta$=1, the entire bed becomes highly charged. (d)  Here, $R_{cg}$ averaged over the entire bed exceeds one, at which point the charge starts to plummet. Data in (a) and (c) obtained by dividing the bed into one-mean-grain-diameter slices and calculating the sum of the absolute values of charges in each slice.}
	\label{heat_map}
\end{figure}

To understand the first two of these observations, we will define a mathematical model that captures the problem's essential dynamics. This will involve some analysis, so we first discuss the simpler asymptotic and breathing states.

We begin with the asymptotic behavior, which provides insight into the bed's charging dynamics. The origin of this behavior can be established by comparing the magnitudes of typical Coulomb and gravitational forces. In Fig.\,\ref{heat_map}, we plot two representative cases, one with moderate charging, ($\chi_e$=0.6, $\eta$=1) and one with rapid charging ($\chi_e$=$\eta$=1). Figures\,\ref{heat_map}(a)-(c) show color-coded charge densities, and Figs.\,\ref{heat_map}(b)-(d) show corresponding bed charges alongside the ratio $R_{cg}$ between characteristic Coulomb and gravitational forces (defined in the figure caption).
 
From Fig.\,\ref{heat_map}(b), we see that for moderate charging, $\bar{q}$ reaches a noisy asymptote after about 5 seconds, at which point $R_{cg}$ averaged over the entire bed approaches 20\%. Figure\,\ref{heat_map}(a), by comparison, shows that the bed charge is dominated by grains in the middle of the bed: we measure that 70\% of the charge is contained in the 14$^{th}$ through 16$^{th}$ layers. If we evaluate $R_{cg}$ in these central layers that dominate bed charging, we find that the charge saturates when $R_{cg}$ reaches one, as shown in Fig.\,\ref{heat_map}(b). We conclude that the charging asymptote coincides with $R_{cg}$ approaching one in the fastest charging region. At this point, grains levitate or stick together - either of which will prevent the collisional charging mechanism that we have described.

As for the breathing state identified in Fig.\,\ref{table_of_exponent}(c) (see also Supplemental video 1), the same behavior occurs, but throughout the entire bed. Figure\,\ref{heat_map}(c) shows that for $\chi_e$=$\eta$=1, where breathing occurs, the highly charged region extends over most of the bed. In this case, Fig.\,\ref{heat_map}(d) shows that $R_{cg}$ averaged over the entire bed reaches one Ð so the whole bed must levitate or stick together.  Indeed, Fig.\,\ref{heat_map}(d) shows that when $R_{cg}$ exceeds one, charging stops, bed charge reduces, and simultaneously the bed contracts. This of course causes densities and collision rates to increase, which in turn must increase charging rates. 

\begin{figure}[htb]
\centering
	\includegraphics[scale=1.5]{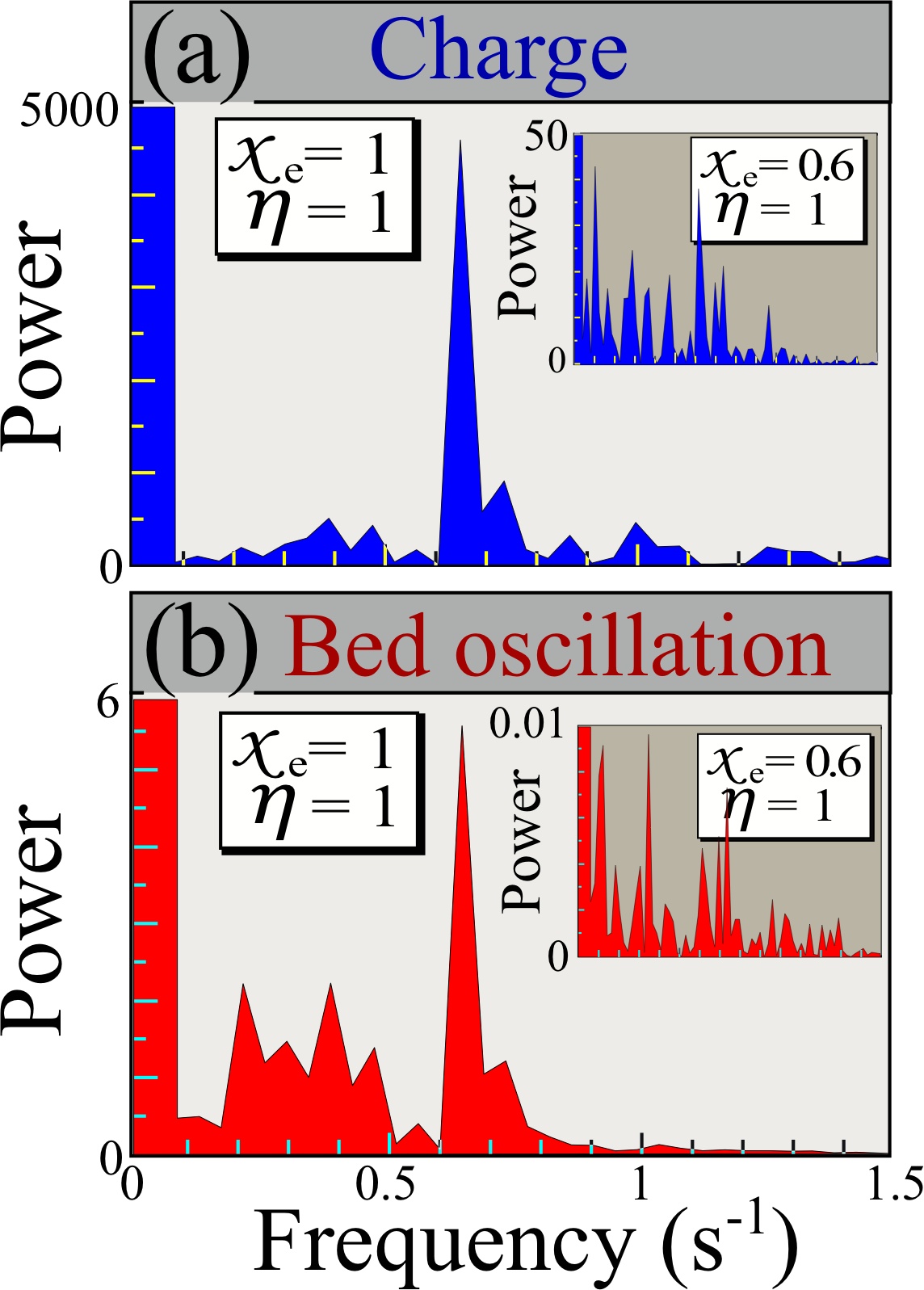}
	\caption{Power spectra of time series obtained by Discrete Fourier Transform. (a) Power spectrum of $\bar{q}(t)$; (b) power spectrum of $\Delta z(t)$. Main plots in (a) \& (b) show breathing state: $\chi_e$=$\eta$=1, and insets show the non-breathing state: $\chi_e$=0.6,  $\eta$=1. Horizontal axes of power spectrum plots are identical.}
	\label{spectrum}
\end{figure}

We confirm the link between charge oscillations and mechanical breathing in Fig.\,\ref{spectrum}, where we compare power spectra of charge and bed expansion in breathing ($\chi_e$=$\eta$=1) and non-breathing ($\chi_e$=0.6, $\eta$=1) states. To evaluate bed oscillations, we average displacements of grain heights, $z_{i}(t)$ from the center of mass height, $z_{c}(t)$: $\Delta z(t) = \langle | z_{c}(t) - z_{i}(t)| \rangle$. As shown in the main plots of  Figs.\,\ref{spectrum}(a)-(b), for the breathing state, both charges and average displacements of grains oscillate at the same frequency, while as shown in the insets, the non-breathing state exhibits broad spectrum noise, with no dominant frequency and much smaller peaks.

Therefore we propose that the cause of both asymptotic charge and breathing oscillations is that the region of the bed that dominates charging reaches $R_{cg}$=1 , at which point particles cannot collide and so cannot charge.  For moderate charging, this occurs over a limited bed height that we presume cannot lift overlying particles; for strong charging, this occurs over the entire bed, which appears to cause global oscillations.

\begin{figure}[htb]
\centering
	\includegraphics[scale=2]{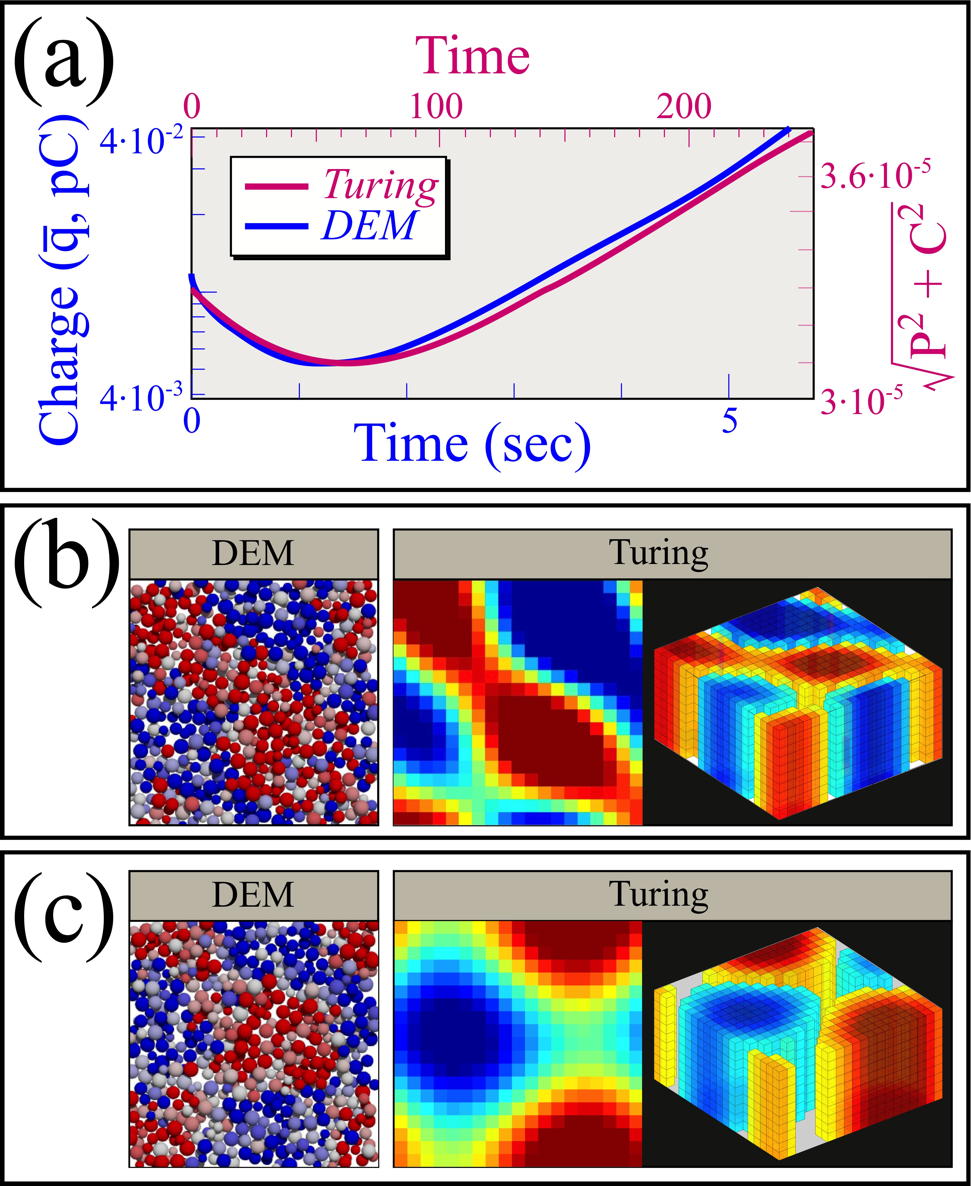}
	\caption{Comparison between DEM and Turing model. (a) Transient from DEM (blue) and Turing simulations (magenta); both ordinate axes are logarithmic; DEM parameters: $\chi_{e}$=0.9, $\eta$= 0.2; Turing parameters: $A$ = 2.25, $A-B$ = -4 (with diffusive terms removed as described in text). (b) Color coded (blue: negative, red: positive) charges in both models using DEM parameters $\chi_{e}$ = 1, $\eta$ = 1, and Turing parameters $A$ = 0.0005 and $A-B$ = -0.0001; DEM shows slice through center of simulation, and Turing results show height average (left) and 3D view (right) of charges. (c) The same views as panel (b), but using DEM parameters $\chi_{e}$ = 1, $\eta$ = 0.6, and Turing parameters $A$ = 0.0005 and $A-B$ = -0.0005. The spotted patterns here are typical, with wavelength that decreases as $A$ and $A-B$ grow.}
	\label{pattern}
\end{figure}

The breathing state indicates that there can be substantial temporal and spatial variations in charging: moreover our DEM simulations also suggest spatial variations in charge density, as shown in Figs.\,\ref{pattern}(b)-(c) (left panels) for two values of $\chi_e$ and $\eta$. In these plots, to identify spatial patterns we enlarge our simulations to $4 \times{10}^{3}$ grains on a horizontal domain of 20$\times$20 mean grain diameters squared.

\subsection*{Mathematical model}
To better understand these spatial variations as well as the transient charge growth, we provide a mathematical model for bed charging and discharging. This model, in the long tradition of dynamical systems analysis, is simplified but captures the essential physics that we have described so far. That physics consists of an iterative growth in polarization, defined by Eq.\,\eqref{charge_induced}, combined with a neutralization in charge, Eq.\,\eqref{charge_transfer}. Accordingly, let us define functions $P(\vec{x}, t)$ and $C(\vec{x},t)$ to represent the polarization in units of polarization per length and charge as functions of position, $\vec{x}$, and time, $t$: 
\begin{align}
\begin{split}
	\frac{dP}{dt}=D_{p} \nabla^{2}P+A(P+C) ,\\
	\frac{dC}{dt}=D_{c} \nabla^{2}C-B(P+C).
\end{split}
\label{reaction_diffusion}
\end{align}
Here, the rate of polarization grows iteratively with electric field in proportion to a single parameter, $A$, and the charge is reduced in proportion to another parameter, $B$. We have included diffusivity terms $D_p$ and $D_c$ to account for migration of the agitated grains, and we anticipate that since polarized grains tend to align and attract one another while charged grains tend to repel, $D_{p}$ should be smaller than $D_{c}$.

We acknowledge that this model is simplified in several respects: it represents polarization as a scalar and neglects nonlinear interactions, particle motion, and Coulomb forces. Nevertheless, as we will show, Eq.\,\eqref{reaction_diffusion} reproduces much of the charging dynamics in the bed shown in Fig.\,\ref{table_of_exponent}, it allows us to cleanly explain the charging transient that we have described, it predicts conditions for the onset of exponential charging, and it permits us to analyze spatial and temporal variations in a transparent and comprehensible way. Moreover, Eq.\,\eqref{reaction_diffusion} is nothing more than the linear reaction-diffusion equation - arguably the simplest of Turing models, whose features have been fully analyzed elsewhere \cite{turing1952chemical}. 

\begin{figure}[htb]
\centering
	\includegraphics[scale=1.5]{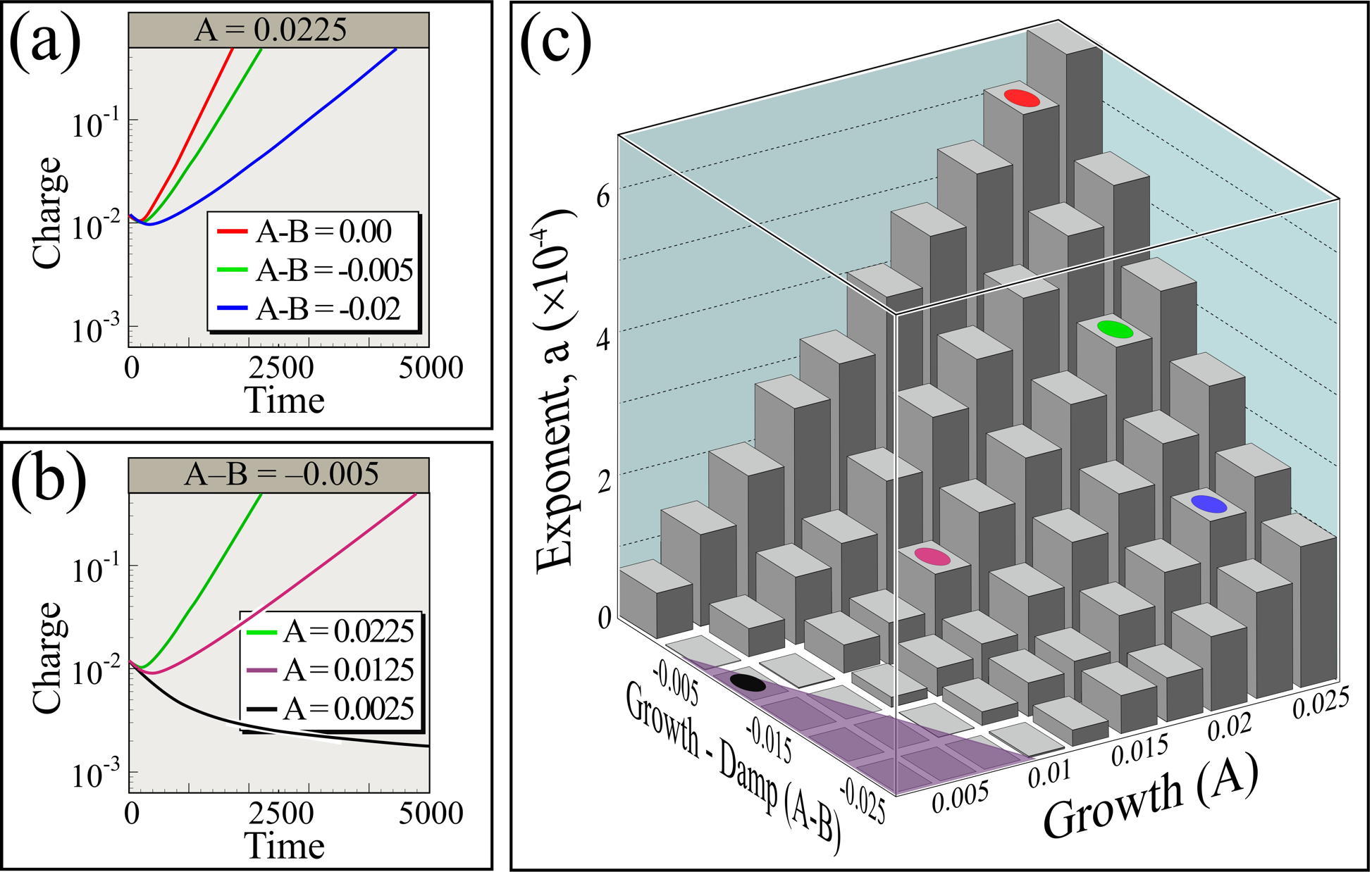}
	\caption{Time evolution of mean charge amplitude, $\bar{C}$, in log-linear plots from solutions of Eq.\,\eqref{reaction_diffusion} obtained through a finite difference integration. As in Fig.\,\ref{table_of_exponent}, polarizations and charges of both signs grow similarly. (a) $A$ is fixed and $A$-$B$ is varied. (b) $A$-$B$ is fixed and $A$ is varied. (c) Exponential growth rates from the slope of a fit of linear regions of log$_{10}$$\bar{C}$ vs. time plots for an array of parameters. Exponents are defined as: $a=\frac{1}{t}log_{10}\left(\frac{\bar{C}(t)}{b}\right)$, where $b$ is a constant. Colored spots correspond to colors used in panels (a) and (b), and violet region should not grow according to stability analysis of Eq.\,\eqref{reaction_diffusion}.}
\label{exponents_rdm}
\end{figure}

We begin our discussion of this mathematical model by reiterating that the parameter $A$ codifies the exponential rate of growth of polarization, as can be seen in the first line of Eq.\,\eqref{reaction_diffusion}. Additionally, considering the equations together, we note that -$B$ causes a reduction in charge, so difference $A$-$B$ determines a net exponential growth. Moreover, neutralization is a physical mechanism that produces charge transfer, which allows particles to gain or lose a net charge. The physical model, Fig.\,\ref{particle}, relies on neutralization to permit charge transfer between particles, and the equivalent term in the Turing model, Eq.\,\eqref{reaction_diffusion}, that allows charge growth is $A-B$. Accordingly, we will associate $A$ with polarization growth - analogous to $\chi_e$ in our DEM simulations - and $A$-$B$ with net growth - analogous to $\eta$.

This correspondence is not exact, but as shown in Fig.\,\ref{exponents_rdm}(c), plotting the exponential growth rates of the charge amplitude averaged over all grains, $\bar{C}$, as a function of $A$-$B$ and $A$ produces substantially similar behavior to that seen in Fig.\,\ref{table_of_exponent}(c). Moreover in detail, if we plot the growth in $\bar{C}$ as a function of time, we obtain panels Figs.\,\ref{exponents_rdm}(a)-(b) for the values indicated. Again the behavior is similar to the growth seen in DEM simulations, Figs.\,\ref{table_of_exponent}(a)-(b). Here we have evolved Eq.\,\eqref{reaction_diffusion} using a  finite difference model integrated with Euler's method with time step 0.5 on a domain of 20$\times$20 horizontal elements by 10 vertical elements. Top and bottom conditions are free and, as in our DEM simulations, horizontal boundary conditions are periodic.

We can go further with Eq.\,\eqref{reaction_diffusion} and evaluate when exponential growth should be seen and determine the cause of the initial transient observed both in DEM and the Turing simulations.

It is well known that the Turing instability (at which point patterned growth emerges) occurs for Eq.\,\eqref{reaction_diffusion} when $A \cdot D_{c}-B\cdot D_{p}>0$ \cite{turing1952chemical}. In Fig.\,\ref{exponents_rdm}, we set $D_{p}=1\cdot 10^{-5}$ and $D_{c} = 4\cdot10^{-5}$, as we mentioned to mimic the fact that charged grains are expected to diffuse faster than polarized ones. So we expect the onset of growing patterns to occur when $A > B/4$. We indicate parameters for which this inequality fails - and so growth is not expected - as a violet shaded region in Fig.\,\ref{exponents_rdm}(c).

So from stability analysis of this model, we predict that the onset of charging in the DEM simulations is associated with a simple Turing mechanism - i.e. that polarization builds faster than diffusion or neutralization can destroy it. On the other hand, parameters that do not generate charging are associated with charge transfer (mediated by neutralization) that is too slow to act before grains diffuse away.

The transient in charging is also amenable to analysis using Eq.\,\eqref{reaction_diffusion}. Our finite difference solutions in Fig.\,\ref{exponents_rdm} are initialized with small random charges (uniformly distributed on [-0.025,0.025]), and since diffusion constants are $O(10^{-5})$, gradients that would trigger diffusion are negligible. If we remove the diffusive terms from Eq.\,\eqref{reaction_diffusion}, we obtain a pair of coupled ordinary differential equations (ODEs) with off-diagonal terms (i.e. $A \cdot C$ in the $\dot P$ equation, and $-B \cdot P$ in the $\dot C$ equation) that are of the same magnitude as the diagonal terms. This is a recipe for non-normal growth \cite{trefethen2005spectra} in which transient contraction can occur in a system whose eigenvalues indicate growth, or vice versa. 

A simple geometrical interpretation of non-normal growth can be obtained by considering a block of clay that is uniformly expanded in $x$ and $y$ (corresponding to positive diagonal terms in a set of ODEs), but is also sheared (corresponding to nonzero off-diagonal terms). In the long term, the expansion will cause points to move away from one another, while in the short term the shearing can temporarily bring points together. 

In the case of Eq.\,\eqref{reaction_diffusion}, non-normal growth occurs when random orientations of vectors ($P$,$C$) re-orient along the expanding eigendirection, approaching smaller values as they do so. This can be confirmed by taking a cluster of points, ($P_i$,$C_i$) near the origin and evolving them according to Eq.\,\eqref{reaction_diffusion} with diffusive terms removed. We use $A$ = 2.25, $B$ = 6.25 for convenience; other values behave similarly. This produces a characteristic transient reduction in $\sum_{i}\sqrt{P_{i}^{2}+C_{i}^{2}}$ followed by growth, as shown in Fig.\,\ref{pattern}(a). Here we evolve 36 points in a grid with $P$ and $C$ between $-10^{-6}$ and $10^{-6}$, and plot $\sum_{i} \sqrt{P_{i}^{2}+C_{i}^{2}}$ to display the transient decrease. The same is seen for either $\sum_{i}|P_{i}|$ or $\sum_{i}|C_{i}|$, separately. In Fig.\,\ref{pattern}(a), we compare this non-normal transient with the transient seen in DEM simulations for the example $\chi_e$ = 0.9, $\eta$ = 0.2.

Apparently the transients shown in Figs.\,\ref{exponents_rdm}(a)-(b), and presumably as well in Figs.\,\ref{table_of_exponent}(a)-(b), are associated with a simple mathematical behavior that occurs while small random vectors align with their ultimate eigendirections. Once alignment occurs, exponential growth can proceed, causing strong local heterogeneities, and at this point, the approximation of neglecting diffusion can no longer be made.

To study these charge heterogeneities, we consider patterns produced both by DEM simulations and by the reaction diffusion model Eq.\,\eqref{reaction_diffusion}, summarized in Fig.\,\ref{pattern}.

In Fig.\,\ref{pattern}(a), we show a comparison between transient growth for the DEM (from Fig.\,\ref{table_of_exponent}(a)) and non-normal growth as described previously. In Fig.\,\ref{pattern}(b), we show a top view of color-coded charges from a DEM simulation that resemble horizontal stripes, along with a comparable Turing pattern from Eq.\,\eqref{reaction_diffusion}. In all of these plots, red indicates positive charge and blue indicates negative. We emphasize, however, that the dominant patterns seen in solutions of the Turing model consist of irregular spots, as shown in Fig.\,\ref{pattern}(c). Alongside each of these simulations, we plot color-coded charges from the Turing simulations in 3D. 

These plots reveal that for aspect ratio 1:2 (i.e. width = 2$\cdot$height), the charge patterns extend throughout the thickness of the bed.  

This occurs for all parameter values shown in Fig.\,\ref{exponents_rdm}, although we note that this is a property that results from the $D_c$ $>$ $D_p$ assumption, which inhibits spatial growth of patterns. As with other reaction-diffusion models, much richer behaviors are also possible: for example relaxing the $D_c$ $>$ $D_p$ constraint leads to temporally oscillatory charges, strong vertical gradients in charge, and other states such as tubes, labyrinths etc \cite{bansagi2011tomography}. Similarly, changing boundary conditions has a strong and well documented \cite{hohenberg1985effects} effect on the patterns expressed - for example modeling a tall thin column of grains rather than a broader bed produces horizontal charge striations \cite{siu2014self}.

Our primary goal for presenting a simplified Turing model has been to show that Eq.\,\eqref{reaction_diffusion} can reproduce charging seen in detailed DEM simulations, as evidenced by comparisons shown in Figs.\,\ref{table_of_exponent}, \,\ref{pattern} and \,\ref{exponents_rdm}, and we defer comprehensive analysis of this model for future studies. The essential thing that the Turing model appears to reveal is that granular charging follows a stereotypical evolution due to straightforward mathematical causes, including a transient decrease in charge followed by exponential growth, a well-defined onset criterion for growth of charging, and emergence of patterned states (chiefly vertical columns of like charge and polarization). The robustness of the mathematics underlying these behaviors suggests that these findings should be common in practical problems, and calls for experimental verification in laboratory and field experiments.

\subsection*{Experiment}
\begin{figure*}[htb]
\centering
	\includegraphics[scale=1.75]{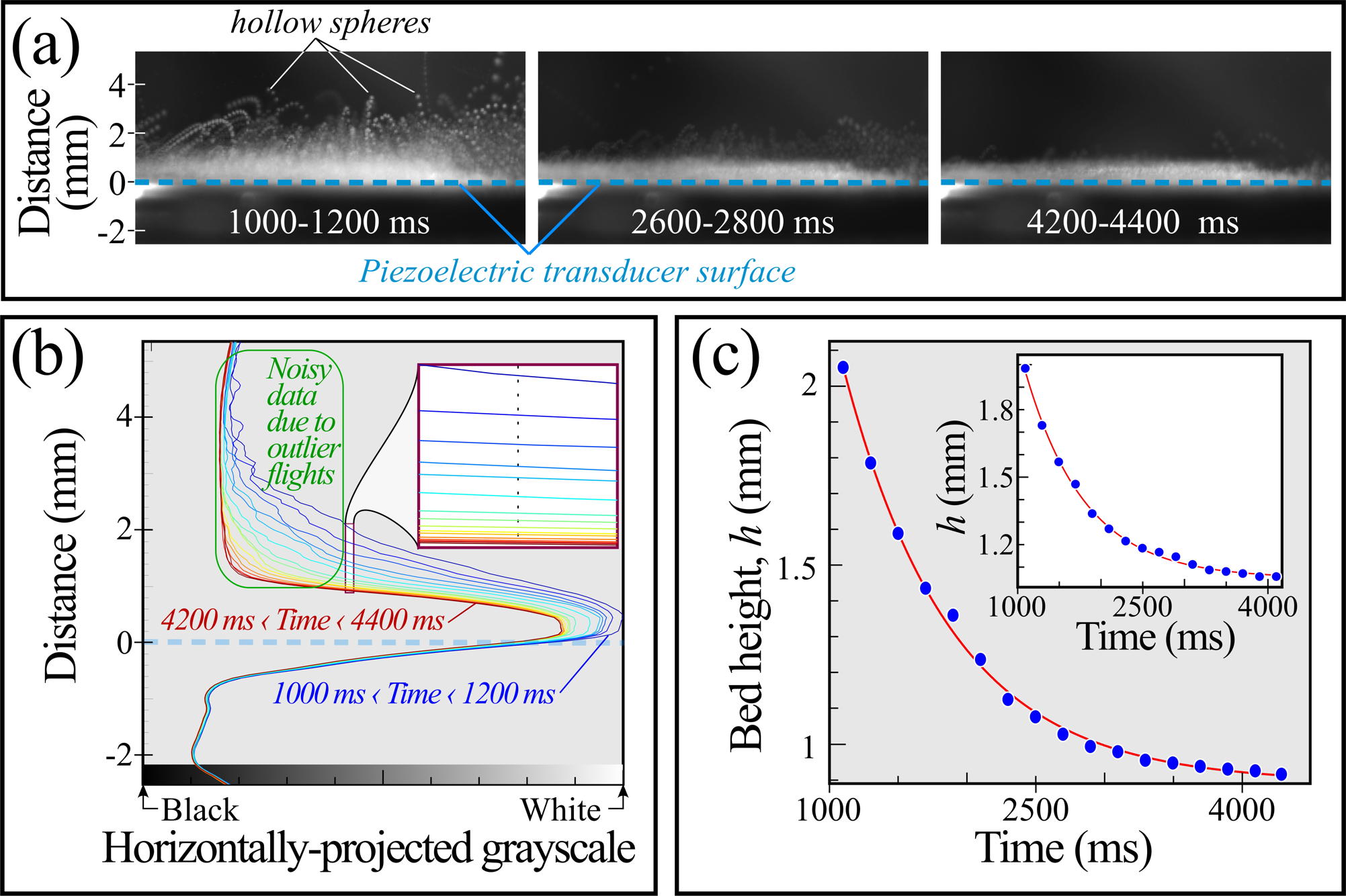}
	\caption{Charging of vibrated hollow glass spheres under microgravity produced in the Bremen drop tower \cite{von2006new}. Gravity is about 10$^{-6}$g and pressure = 1 mbar. Spheres have density 0.14 g/cm$^3$ and diameters between 125 and 150 $\mu$m (Cospheric LLC, Santa Barbara, CA). (a) Typical time-lapse images showing 200 ms superpositions of video frames taken at 110 frames per second. At time = 0, vibration of the transducer shown is initiated, after which the apparatus is rapidly accelerated by catapult, and by several tens of milliseconds, microgravity is achieved. Gravity is nearly nonexistent, so parabolic trajectories can only be due to electrostatics. (b) Horizontal projections of grayscale of time-lapse images are used to evaluate height of bed of agitated particles. Inset shows enlargement just beyond noisy region associated with irregular large particle flights; intersections with dotted line used to estimate bed heights. (c) Estimates of bed height obtained from intersections shown in panel (b), along with exponential fit. Inset shows second experiment where metal plate is covered by insulating tape.  
}
\label{experiment}
\end{figure*}

The underlying hypothesis of both DEM and Turing models is that an iterative process leads to exponentially rapid growth of initially infinitesimally small charges, so we close by testing this hypothesis in experiments. In these experiments, shown in Fig.\,\ref{experiment}, hollow glass spheres are vibrated on a grounded metal plate at 2 kHz by a piezoelectric transducer. The thickness of the particle bed is close to that used in our DEM simulations (under 1 mm, or about 9 particle diameters). These experiments are performed under microgravity (see figure caption), yet as shown in Fig.\,\ref{experiment}(a), particles return to the plate along curved trajectories. Crucially, since gravity is essentially absent, the only known force that can act at a distance in this way is electrostatic. Moreover, the heights of particle flights diminish with time as can be seen in Fig.\,\ref{experiment}(a) and in Supplemental video 2, which we can use to obtain a quantitative evaluation of our DEM and Turing predictions, as follows.

The maximum height, $h$, of a particle ballistically ejected from the bed is simply its kinetic energy, $KE$, divided by the force, $F$, attracting the particle to the bed: $h \sim{KE/F}$. We have seen that our model predicts exponential growth in charges of both signs, so the force, $F$, associated with these charges must also grow exponentially in time. Since $h \sim{1/F}$, we predict that $h$ will decrease exponentially in time: $h \sim e^{-a \cdot time}$ where $a$ defines the charging rate shown in Figs.\,\ref{table_of_exponent} and \,\ref{exponents_rdm}.

We assess this prediction by evaluating heights reached by particles near the top of the bed. As shown in Fig.\,\ref{experiment}(b), we horizontally sum the grayscales of pixels from successive 200 ms superpositions (as in Fig.\,\ref{experiment}(a)). The transducer could drift away from the agitated particle bed and change the particle number density that would affect the brightness of the images. However, the timescales for particle losses qualitatively are on the order of 10s while the timescales for charging are a fraction of a second. Keeping these two different timescales in mind, we consider the brightness as a suitable measure of the charging.
  
Our procedure is as follows. High flying outlier particles produce noisy variations in grayscale, so we exclude the noisy region identified in Fig.\,\ref{experiment}(b), and select a moderate grayscale that shows little noise but provides the largest available height discrimination between superpositions. This grayscale is boxed in the main plot and enlarged in the inset to this figure. We evaluate the grayscale at the center of this region (broken line in the inset), which we plot in Fig.\,\ref{experiment}(c), along with a least-squares fit to the predicted exponential, $h=h_{0}+h_{1}e^{-a \cdot time}$. We find that a fit can be made using $a = 1.31 \pm 0.03$ sec$^{-1}$ with correlation coefficient, $r^2$=0.997. We repeat the experiment with the metal plate covered with insulating tape, and obtain the height vs. time plot shown in the inset to Fig.\,\ref{experiment}(c): here we obtain $a = 1.40 \pm 0.03$ sec$^{-1}$ and $r^2$=0.998. Both of these fits have growth rates, $a$, in the range expected from Fig.\,\ref{table_of_exponent}(c).
   
We are aware that other dependencies can also be fitted. However, this would require an alternative competitive idea to be tested, which does not currently exist. Consequently we find that the experimental results are consistent with the theoretical prediction and avoid drawing conclusions about the functional dependence just from the analysis of the experimental data.
 
These results seem to confirm our predictions of exponential charging, however other possibilities deserve mention. First, most bouncing particles return to the bed, yet some particles near the edges escape (see Supplemental video 2). It might be argued that loss of particles could account for the decrease in bed height, however we note that fewer bed particles would cause particles above the bed both to be less strongly attracted to the bed and to rebound more elastically, both of which would increase, rather than decrease, the measured heights shown in Fig.\,\ref{experiment}.
 
Second, it is possible that particles have been tribocharged by the vibrating plate. Although tribocharging doubtless occurs, we remark that (1) particles ejected from the bed are attracted back to the bed, so particles cannot simply be tribocharged with the same sign, which would cause repulsion; (2) spheres landing on grounded metal and on insulating tape produce nearly indistinguishable results; and (3) charging appears to occur exponentially in time. None of these results are consistent with tribocharging as it is traditionally understood \cite{cimarelli2014experimental}. It remains conceivable that particles near and far from the vibrating plate could acquire opposite charges \cite{kolehmainen2016hybrid}, however this would not explain the apparent exponential charging.

Third, several groups have described charging models for particles differing in size \cite{forward2009charge, kok2009electrification, waitukaitis2014size}, and indeed our hollow spheres range from 125 $\mu$m to 150 $\mu$m. Again, size-dependent charging doubtless does occur, however such mechanisms invariably produce monotonically decreasing charging rates, rather than the exponential growth that we observe.

Based on these considerations, we conclude that our agitated granular bed appears to produce exponential growth in charges of both signs, which to our knowledge our model is unique in predicting.

\section*{\label{discussion}Discussion}
We have performed simulations, modeling, and experiments of charged grains in an agitated bed. The simulations show that grains can charge exponentially rapidly by feeding back their electric fields through their neighbors. The Turing model provides a simple framework to understand the exponential growth in polarization and charge as well as more detailed predictions such as an onset criterion and non-normal charging transients. The microgravity experiments confirm that charging of agitated beds of insulating grains does appear to grow exponentially. Finally, the simulations and the Turing model predict a previously unreported oscillatory state and complex spatiotemporal charging dynamics, both of which merit further study.

We propose that our findings of exponential growth of charging may account for the generation of multi-million volt potentials observed in nature, and may contribute to improved understanding of electrical charging in mining \cite{hertzberg1987introduction}, and industrial powder handling \cite{mehrotra2007spontaneous, schein2007recent, kolehmainen2016hybrid}.

Beyond these findings, we remark that a Turing model contains a natural mechanism for producing ``something from nothing," which we mentioned at the start characterizes the emergence of strong electrical effects by doing no more than agitating initially nearly neutral grains. Indeed, historically the appearance of complex dynamics from seemingly benign reactions was rejected as being ``impossible \cite{winfree1984prehistory}" for this reason. In our problem, the ``reaction" comes in the form of known electrical effects, polarization and neutralization, but at its core these are not mathematically different from reactions between enzymes or autocatalysts - which also appear to produce something from nothing.

\appendix*
\section*{\label{methods}Methods}
\begin{figure}[htb]
\centering
\includegraphics[width=9cm]{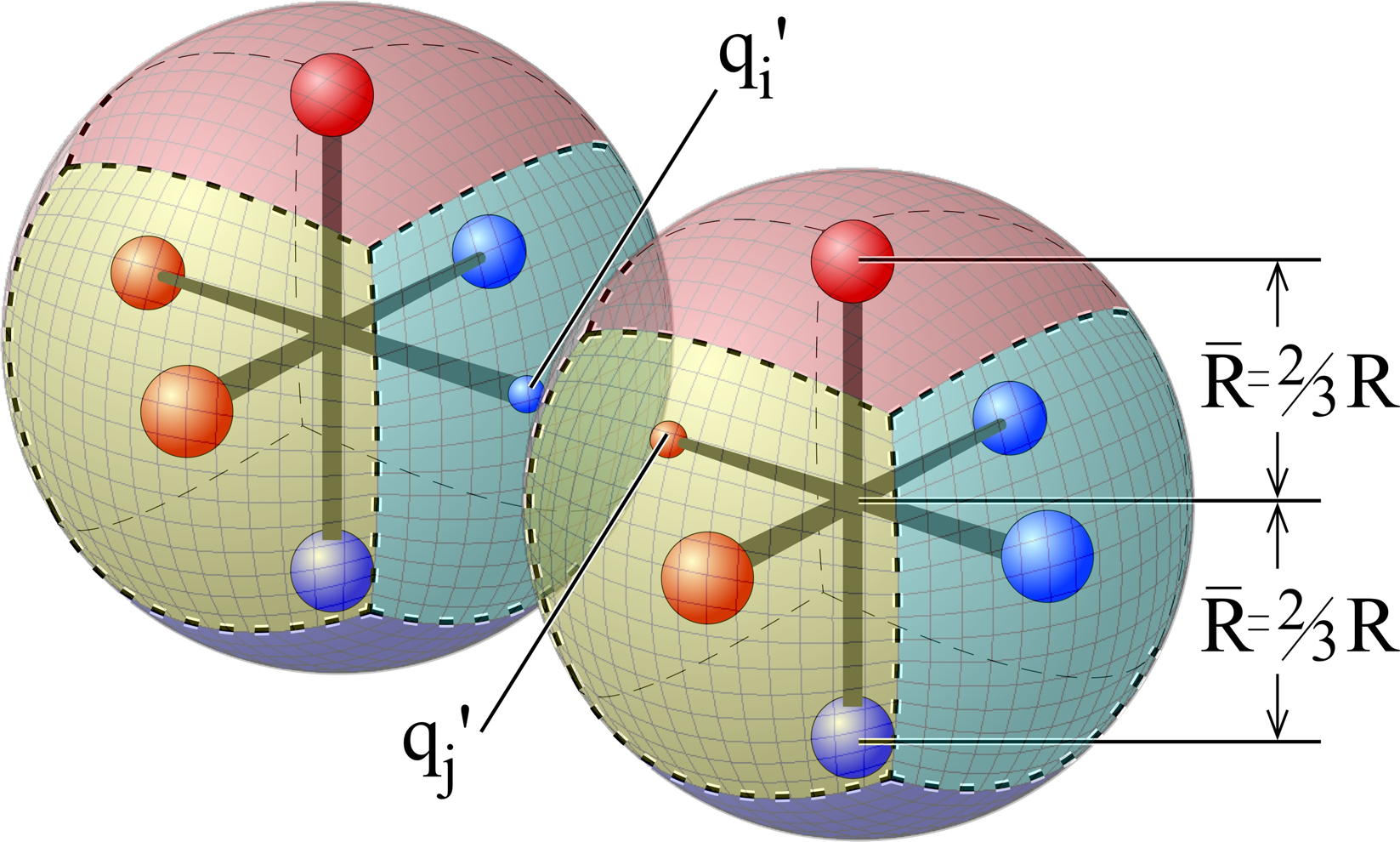}
\caption{Schematic representation of grains with six independent charge domains immediately following the neutralization process. Each charge domain is fixed at 2/3 of the mean grain radius from the center of the grain. Two contacting domain charges, $q_i$ and $q_j$, neutralize at a contact point to become $q'_i$ and $q'_j$. Size of the domain charges here represent the magnitude of the charges.}
\label{particle}
\end{figure}
To account for a heterogeneous permanent charge distribution on the surface of insulating grains, we considered a multipole expansion and truncated after the second-order term. We simulate this by embedding in each grain six independent, orthogonally placed charge domains as shown in Fig.\,\ref{particle}. As in previous work \cite{C6SM00357E}, we track translational and rotational motions of each grain by evaluating its dynamics due to mechanical and electrical forces and torques acting on it. We then solve the equations of motion for each grain by means of the DEM on a domain that is periodic along the horizontal directions. We achieve periodicity by surrounding the computational domain by 8 copies in the horizontal plane, which we use for field and energy calculations.

The DEM algorithm itself uses spherical grains with restitution coefficient 0.935 and kinetic friction coefficient of 0.4. We use the model of Walton and Braun \cite{walton1986viscosity} for the elastic force and fix the two elastic coefficients, $k_l$=0.07 and $k_u$=0.08, to achieve the restitution coefficient, $\sqrt{k_{l}/k_{u}}=\sqrt{0.07/0.08} \approx 0.935$. We use a time step of 50 msec., which produces over 10$^{2}$ steps per collision for the fastest moving grains. We use polydisperse grain sizes to prevent crystallization: the radius of each grain, $R_i$, is Gaussian distributed with standard deviation 10\% of the mean radius, $\bar{R}=0.75$ mm, and each grain has the density of glass, $\rho_{g}$ = 2.4 g/cm$^3$. 

The top of the computational domain is free, and the bottom is fixed. Any grain that hits the bottom acquires additional upward energy defined by a kick velocity, $\vec{V}=2.7\sqrt{2g\bar{R}}\hat{z}$, which maintains the granular bed in a collisional state.

Particles polarize according to Eq.\,\eqref{charge_induced}, where we emphasize that $\vec{E}_{i}$ is the electric field at the center of the $i^{th}$ grain due to all pre-existing permanent charges in the system. The distinction between permanent and induced charges is significant because induced charges are slaved to the external field, and cannot themselves do work. For example, induced charge dipoles always point in the direction of an external field and so cannot exert torque on a grain. Permanent charges, on the contrary, are fixed on a grain and exert forces on other charges \cite{lee2015direct}. An induced polarization, given as a superposition of multiple orders of electrostatic interaction, could point at any arbitrary direction. The effects of higher order polarizations, even for particles in close proximity, are appropriately captured in this scheme. It is only when particles collide that charges resulting from induced dipoles are mapped onto the domains to become permanent (described next).

At the instant when two grains collide, neutralization is imposed between contacting charge domains of colliding grains $i$ and $j$ according to Eq.\,\eqref{charge_transfer}. This permits charge transfer between grains, so for $\eta$=0 all charges remain unchanged, and grains increasingly transfer charges as $\eta$ grows. Equation \,\eqref{charge_transfer} is applied during binary collisions, and whenever a grain contacts multiple neighbors during a single time step, we perform this operation for all pairs in random order.
 
Both permanent and induced charges take part in neutralization events, and to keep accounts straight we add exactly the fraction of induced charges needed to conserve charge to the permanent charges. That is, if an induced charge $\Delta q$ is added to one domain of a grain due to Eq.\,\eqref{charge_transfer}, then -$\Delta q$ will be made permanent on the opposing domain of the same grain to guarantee overall charge neutrality. Finally, we prevent spurious repetition of charging by only applying neutralization and induction operations at the moment when two grains first touch one another. 

We enforce energy conservation in two ways. First, we compensate for the energy associated with assembling the induced dipole moment prescribed by Eq.\,\eqref{charge_induced} by integrating the work needed to bring the induced charges to their positions from infinity \cite{griffiths1999introduction}. We then evaluate the gradient of this energy, which gives us a mechanical force that we apply to each particle. This is the force that must be exerted to polarize the particle, and the spatial integral of this gradient is mechanical work that exactly equals the required electrical energy.
  
Second, we note that an induced dipole moment changes the electric field of neighboring grains, and this change in turn induces secondary dipole moments according to Eq.\,\eqref{charge_induced}. To conserve energy, we account for secondary dipoles by reducing each primary moment by exactly the energy associated with every secondary moment. This process feeds back iteratively, so that every secondary moment in turn induces another moment on the originating dipole. We have numerically confirmed that this feedback converges rapidly, and after two iterations the error in neglecting higher order terms is less than 0.8\%. Consequently, in our simulations we perform two iterations of inducing new additive dipole moments based on this feedback process.
  
Neutralization events also involve energy considerations: when a particle of charge $q_i$ fully neutralizes ($\eta$=1) during collision with a neighbor, each particle will leave the collision with a charge of up to $q_i/2$. This produces repulsion between the particles that was not present prior to the collision. This repulsion is a real physical effect that is seen in experiments \cite{shinbrot2006triboelectrification}, and so we include the repulsion in our simulations.

We initialize each simulation by dropping 10$^{3}$ grains onto the fixed bottom, of area 10$\bar{R}\times$10$\bar{R}$. No energy is injected (through kicks by the bottom surface) while particles settle, and grains are all initially neutral. We wait five seconds until grain velocities become negligibly small (kick velocity/10$^{3}$).  We then add charges uniformly distributed on [-10$^{-2}$, 10$^{-2}$] pC to all six domains of all grains and thereafter kick particles at the bottom surface.

\bibliographystyle{naturemag}

%\bibliography{bibliography}

\begin{thebibliography}{}
\expandafter\ifx\csname url\endcsname\relax
  \def\url#1{\texttt{#1}}\fi
\expandafter\ifx\csname urlprefix\endcsname\relax\def\urlprefix{URL }\fi
\providecommand{\bibinfo}[2]{#2}
\providecommand{\eprint}[2][]{\url{#2}}

\end{thebibliography}


\begin{thebibliography}{10}
\expandafter\ifx\csname url\endcsname\relax
  \def\url#1{\texttt{#1}}\fi
\expandafter\ifx\csname urlprefix\endcsname\relax\def\urlprefix{URL }\fi
\providecommand{\bibinfo}[2]{#2}
\providecommand{\eprint}[2][]{\url{#2}}

\bibitem{anderson1965electricity}
\bibinfo{author}{Anderson, R.} \emph{et~al.}
\newblock \bibinfo{title}{Electricity in volcanic clouds}.
\newblock \emph{\bibinfo{journal}{Science}} \textbf{\bibinfo{volume}{148}},
  \bibinfo{pages}{1179--1189} (\bibinfo{year}{1965}).

\bibitem{shaw1929tribo}
\bibinfo{author}{Shaw, P.}
\newblock \bibinfo{title}{Tribo-electricity and friction. iv. electricity due
  to air-blown particles}.
\newblock \emph{\bibinfo{journal}{Proceedings of the Royal Society of London.
  Series A, Containing Papers of a Mathematical and Physical Character}}
  \textbf{\bibinfo{volume}{122}}, \bibinfo{pages}{49--58}
  (\bibinfo{year}{1929}).

\bibitem{mehrotra2007spontaneous}
\bibinfo{author}{Mehrotra, A.}, \bibinfo{author}{Muzzio, F.~J.} \&
  \bibinfo{author}{Shinbrot, T.}
\newblock \bibinfo{title}{Spontaneous separation of charged grains}.
\newblock \emph{\bibinfo{journal}{Physical review letters}}
  \textbf{\bibinfo{volume}{99}}, \bibinfo{pages}{058001}
  (\bibinfo{year}{2007}).

\bibitem{hertzberg1987introduction}
\bibinfo{author}{Hertzberg, M.} \& \bibinfo{author}{Cashdollar, K.~L.}
\newblock \bibinfo{title}{Introduction to dust explosions}.
\newblock In \emph{\bibinfo{booktitle}{Industrial dust explosions}}
  (\bibinfo{publisher}{ASTM International}, \bibinfo{year}{1987}).

\bibitem{schein2007recent}
\bibinfo{author}{Schein, L.}
\newblock \bibinfo{title}{Recent progress and continuing puzzles in
  electrostatics}.
\newblock \emph{\bibinfo{journal}{Science}} \textbf{\bibinfo{volume}{316}},
  \bibinfo{pages}{1572--1573} (\bibinfo{year}{2007}).

\bibitem{lowell1986triboelectrification}
\bibinfo{author}{Lowell, J.} \& \bibinfo{author}{Truscott, W.}
\newblock \bibinfo{title}{Triboelectrification of identical insulators. ii.
  theory and further experiments}.
\newblock \emph{\bibinfo{journal}{Journal of Physics D: Applied Physics}}
  \textbf{\bibinfo{volume}{19}}, \bibinfo{pages}{1281} (\bibinfo{year}{1986}).

\bibitem{terris1989contact}
\bibinfo{author}{Terris, B.}, \bibinfo{author}{Stern, J.},
  \bibinfo{author}{Rugar, D.} \& \bibinfo{author}{Mamin, H.}
\newblock \bibinfo{title}{Contact electrification using force microscopy}.
\newblock \emph{\bibinfo{journal}{Physical Review Letters}}
  \textbf{\bibinfo{volume}{63}}, \bibinfo{pages}{2669} (\bibinfo{year}{1989}).

\bibitem{forward2009charge}
\bibinfo{author}{Forward, K.~M.}, \bibinfo{author}{Lacks, D.~J.} \&
  \bibinfo{author}{Sankaran, R.~M.}
\newblock \bibinfo{title}{Charge segregation depends on particle size in
  triboelectrically charged granular materials}.
\newblock \emph{\bibinfo{journal}{Physical review letters}}
  \textbf{\bibinfo{volume}{102}}, \bibinfo{pages}{028001}
  (\bibinfo{year}{2009}).

\bibitem{kok2009electrification}
\bibinfo{author}{Kok, J.~F.} \& \bibinfo{author}{Lacks, D.~J.}
\newblock \bibinfo{title}{Electrification of granular systems of identical
  insulators}.
\newblock \emph{\bibinfo{journal}{Physical Review E}}
  \textbf{\bibinfo{volume}{79}}, \bibinfo{pages}{051304}
  (\bibinfo{year}{2009}).

\bibitem{waitukaitis2014size}
\bibinfo{author}{Waitukaitis, S.~R.}, \bibinfo{author}{Lee, V.},
  \bibinfo{author}{Pierson, J.~M.}, \bibinfo{author}{Forman, S.~L.} \&
  \bibinfo{author}{Jaeger, H.~M.}
\newblock \bibinfo{title}{Size-dependent same-material tribocharging in
  insulating grains}.
\newblock \emph{\bibinfo{journal}{Physical Review Letters}}
  \textbf{\bibinfo{volume}{112}}, \bibinfo{pages}{218001}
  (\bibinfo{year}{2014}).

\bibitem{cademartiri2012simple}
\bibinfo{author}{Cademartiri, R.} \emph{et~al.}
\newblock \bibinfo{title}{A simple two-dimensional model system to study
  electrostatic-self-assembly}.
\newblock \emph{\bibinfo{journal}{Soft Matter}} \textbf{\bibinfo{volume}{8}},
  \bibinfo{pages}{9771--9791} (\bibinfo{year}{2012}).

\bibitem{kolehmainen2016hybrid}
\bibinfo{author}{Kolehmainen, J.}, \bibinfo{author}{Ozel, A.},
  \bibinfo{author}{Boyce, C.~M.} \& \bibinfo{author}{Sundaresan, S.}
\newblock \bibinfo{title}{A hybrid approach to computing electrostatic forces
  in fluidized beds of charged particles}.
\newblock \emph{\bibinfo{journal}{AIChE Journal}}
  \textbf{\bibinfo{volume}{62}}, \bibinfo{pages}{2282--2295}
  (\bibinfo{year}{2016}).

\bibitem{harper1998contact}
\bibinfo{author}{Harper, W.~R.}
\newblock \emph{\bibinfo{title}{Contact and frictional electrification}}
  (\bibinfo{publisher}{Laplacian Press Morgan Hill, CA}, \bibinfo{year}{1998}).

\bibitem{shinbrot2008spontaneous}
\bibinfo{author}{Shinbrot, T.}, \bibinfo{author}{Komatsu, T.} \&
  \bibinfo{author}{Zhao, Q.}
\newblock \bibinfo{title}{Spontaneous tribocharging of similar materials}.
\newblock \emph{\bibinfo{journal}{EPL (Europhysics Letters)}}
  \textbf{\bibinfo{volume}{83}}, \bibinfo{pages}{24004} (\bibinfo{year}{2008}).

\bibitem{baytekin2011mosaic}
\bibinfo{author}{Baytekin, H.} \emph{et~al.}
\newblock \bibinfo{title}{The mosaic of surface charge in contact
  electrification}.
\newblock \emph{\bibinfo{journal}{Science}} \textbf{\bibinfo{volume}{333}},
  \bibinfo{pages}{308--312} (\bibinfo{year}{2011}).

\bibitem{pahtz2010particle}
\bibinfo{author}{P{\"a}htz, T.}, \bibinfo{author}{Herrmann, H.} \&
  \bibinfo{author}{Shinbrot, T.}
\newblock \bibinfo{title}{Why do particle clouds generate electric charges?}
\newblock \emph{\bibinfo{journal}{Nature Physics}}
  \textbf{\bibinfo{volume}{6}}, \bibinfo{pages}{364--368}
  (\bibinfo{year}{2010}).

\bibitem{C6SM00357E}
\bibinfo{author}{Yoshimatsu, R.}, \bibinfo{author}{Araujo, N. A.~M.},
  \bibinfo{author}{Shinbrot, T.} \& \bibinfo{author}{Herrmann, H.}
\newblock \bibinfo{title}{Field driven charging dynamics of a fluidized
  granular bed}.
\newblock \emph{\bibinfo{journal}{Soft Matter}} \textbf{\bibinfo{volume}{12}},
  \bibinfo{pages}{6261--6267} (\bibinfo{year}{2016}).

\bibitem{zheng2004theoretical}
\bibinfo{author}{Zheng, X.}, \bibinfo{author}{He, L.} \& \bibinfo{author}{Zhou,
  Y.}
\newblock \bibinfo{title}{Theoretical model of the electric field produced by
  charged particles in windblown sand flux}.
\newblock \emph{\bibinfo{journal}{Journal of Geophysical Research: Atmospheres
  (1984--2012)}} \textbf{\bibinfo{volume}{109}} (\bibinfo{year}{2004}).

\bibitem{rasmussen2009enhancement}
\bibinfo{author}{Rasmussen, K.~R.}, \bibinfo{author}{Kok, J.~F.} \&
  \bibinfo{author}{Merrison, J.~P.}
\newblock \bibinfo{title}{Enhancement in wind-driven sand transport by electric
  fields}.
\newblock \emph{\bibinfo{journal}{Planetary and Space Science}}
  \textbf{\bibinfo{volume}{57}}, \bibinfo{pages}{804--808}
  (\bibinfo{year}{2009}).

\bibitem{zhang2015electric}
\bibinfo{author}{Zhang, Y.} \emph{et~al.}
\newblock \bibinfo{title}{Electric field and humidity trigger contact
  electrification}.
\newblock \emph{\bibinfo{journal}{Physical Review X}}
  \textbf{\bibinfo{volume}{5}}, \bibinfo{pages}{011002} (\bibinfo{year}{2015}).

\bibitem{siu2014self}
\bibinfo{author}{Siu, T.}, \bibinfo{author}{Cotton, J.},
  \bibinfo{author}{Mattson, G.} \& \bibinfo{author}{Shinbrot, T.}
\newblock \bibinfo{title}{Self-sustaining charging of identical colliding
  particles}.
\newblock \emph{\bibinfo{journal}{Physical Review E}}
  \textbf{\bibinfo{volume}{89}}, \bibinfo{pages}{052208}
  (\bibinfo{year}{2014}).

\bibitem{granular}
\bibinfo{author}{T.Poschel} \& \bibinfo{author}{Schwager, T.}
\newblock \emph{\bibinfo{title}{Computational Granular Dynamics}}
  (\bibinfo{publisher}{Springer Berlin Heidelberg}, \bibinfo{year}{2005}).

\bibitem{turing1952chemical}
\bibinfo{author}{Turing, A.~M.}
\newblock \bibinfo{title}{The chemical basis of morphogenesis}.
\newblock \emph{\bibinfo{journal}{Philosophical Transactions of the Royal
  Society of London B: Biological Sciences}} \textbf{\bibinfo{volume}{237}},
  \bibinfo{pages}{37--72} (\bibinfo{year}{1952}).

\bibitem{trefethen2005spectra}
\bibinfo{author}{Trefethen, L.~N.} \& \bibinfo{author}{Embree, M.}
\newblock \emph{\bibinfo{title}{Spectra and pseudospectra: the behavior of
  nonnormal matrices and operators}} (\bibinfo{publisher}{Princeton University
  Press}, \bibinfo{year}{2005}).

\bibitem{bansagi2011tomography}
\bibinfo{author}{B{\'a}ns{\'a}gi, T.}, \bibinfo{author}{Vanag, V.~K.} \&
  \bibinfo{author}{Epstein, I.~R.}
\newblock \bibinfo{title}{Tomography of reaction-diffusion microemulsions
  reveals three-dimensional turing patterns}.
\newblock \emph{\bibinfo{journal}{Science}} \textbf{\bibinfo{volume}{331}},
  \bibinfo{pages}{1309--1312} (\bibinfo{year}{2011}).

\bibitem{hohenberg1985effects}
\bibinfo{author}{Hohenberg, P.}, \bibinfo{author}{Kramer, L.} \&
  \bibinfo{author}{Riecke, H.}
\newblock \bibinfo{title}{Effects of boundaries on one-dimensional
  reaction-diffusion equations near threshold}.
\newblock \emph{\bibinfo{journal}{Physica D: Nonlinear Phenomena}}
  \textbf{\bibinfo{volume}{15}}, \bibinfo{pages}{402--420}
  (\bibinfo{year}{1985}).

\bibitem{cimarelli2014experimental}
\bibinfo{author}{Cimarelli, C.}, \bibinfo{author}{Alatorre-Ibarg{\"u}engoitia,
  M.}, \bibinfo{author}{Kueppers, U.}, \bibinfo{author}{Scheu, B.} \&
  \bibinfo{author}{Dingwell, D.~B.}
\newblock \bibinfo{title}{Experimental generation of volcanic lightning}.
\newblock \emph{\bibinfo{journal}{Geology}} \textbf{\bibinfo{volume}{42}},
  \bibinfo{pages}{79--82} (\bibinfo{year}{2014}).

\bibitem{winfree1984prehistory}
\bibinfo{author}{Winfree, A.~T.}
\newblock \bibinfo{title}{The prehistory of the belousov-zhabotinsky
  oscillator}.
\newblock \emph{\bibinfo{journal}{J. Chem. Educ}}
  \textbf{\bibinfo{volume}{61}}, \bibinfo{pages}{661} (\bibinfo{year}{1984}).

\bibitem{walton1986viscosity}
\bibinfo{author}{Walton, O.~R.} \& \bibinfo{author}{Braun, R.~L.}
\newblock \bibinfo{title}{Viscosity, granular-temperature, and stress
  calculations for shearing assemblies of inelastic, frictional disks}.
\newblock \emph{\bibinfo{journal}{Journal of Rheology (1978-present)}}
  \textbf{\bibinfo{volume}{30}}, \bibinfo{pages}{949--980}
  (\bibinfo{year}{1986}).

\bibitem{lee2015direct}
\bibinfo{author}{Lee, V.}, \bibinfo{author}{Waitukaitis, S.~R.},
  \bibinfo{author}{Miskin, M.~Z.} \& \bibinfo{author}{Jaeger, H.~M.}
\newblock \bibinfo{title}{Direct observation of particle interactions and
  clustering in charged granular streams}.
\newblock \emph{\bibinfo{journal}{Nature Physics}}  (\bibinfo{year}{2015}).

\bibitem{griffiths1999introduction}
\bibinfo{author}{Griffiths, D.~J.} \& \bibinfo{author}{College, R.}
\newblock \emph{\bibinfo{title}{Introduction to electrodynamics}},
  vol.~\bibinfo{volume}{3} (\bibinfo{publisher}{prentice Hall Upper Saddle
  River, NJ}, \bibinfo{year}{1999}).

\bibitem{shinbrot2006triboelectrification}
\bibinfo{author}{Shinbrot, T.}, \bibinfo{author}{LaMarche, K.} \&
  \bibinfo{author}{Glasser, B.~J.}
\newblock \bibinfo{title}{Triboelectrification and razorbacks: geophysical
  patterns produced in dry grains}.
\newblock \emph{\bibinfo{journal}{Physical review letters}}
  \textbf{\bibinfo{volume}{96}}, \bibinfo{pages}{178002}
  (\bibinfo{year}{2006}).

\bibitem{von2006new}
\bibinfo{author}{Von~Kampen, P.}, \bibinfo{author}{Kaczmarczik, U.} \&
  \bibinfo{author}{Rath, H.~J.}
\newblock \bibinfo{title}{The new drop tower catapult system}.
\newblock \emph{\bibinfo{journal}{Acta Astronautica}}
  \textbf{\bibinfo{volume}{59}}, \bibinfo{pages}{278--283}
  (\bibinfo{year}{2006}).

\end{thebibliography}

\section*{Acknowledgments}
We acknowledge financial support from the ETH Grant, the ETH Risk Center, the Brazilian institute INCT-SC, and the European Research Council (ERC) Advanced Grant 319968-FlowCCS. NA acknowledges financial support from the Portuguese Foundation for Science and Technology (FCT) under Contracts nos. EXCL/FIS-NAN/0083/2012, UID/FIS/00618/2013, and IF/00255/2013. The experiments are supported by the DLR Space Management with funds provided by the Federal Ministry ofÊEconomics and Technology (BMWi) under Grant No. DLR 50 WM 1542. TS acknowledges support from the NSF DMR, award $\sharp$1404792.

\section*{Author contributions}
RY, NA, GW, HH, and TS  have contributed equally in carrying out the simulations, designing the experiment, analyzing the results, and writing the manuscript.
  
\section*{Competing financial interests}
The authors declare no competing financial interests.
\clearpage
\end{document}